\documentclass[pra,aps,amsfonts,amsmath,twocolumn,showpacs,floatfix]{revtex4}
\usepackage{bm}
\usepackage{amsfonts}
\usepackage{amsmath}
\usepackage{graphicx}
\newcommand{\Eq}[1]{Eq.\@ (\ref{#1})}
\newcommand{\Eqs}[1]{Eqs.\@ (\ref{#1})}
\newcommand{\Refe}[1]{Ref.\@ \cite{#1}}
\newcommand{\Fig}[1]{Fig.\@ \ref{#1}}

\newcommand{\Sec}[1]{Sec.\@ \ref{#1}}
\DeclareMathOperator{\Real}{Re}
\DeclareMathOperator{\Imag}{Im}

\DeclareMathOperator{\sgn}{sgn}
\DeclareMathOperator{\arctanh}{arctanh}
\renewcommand{\Re}{\Real}
\renewcommand{\Im}{\Imag}
\newcommand{\poisson}[2]{\{#1,#2\}}

\newcommand{\jv}{\bm{\mathrm{j}}}
\newcommand{\kv}{\bm{\mathrm{k}}}
\newcommand{\nablav}{\bm{\nabla}}
\newcommand{\pv}{\bm{\mathrm{p}}}
\newcommand{\Pv}{\bm{\mathrm{P}}}
\newcommand{\rv}{\bm{\mathrm{r}}}
\newcommand{\Rv}{\bm{\mathrm{R}}}
\newcommand{\sv}{\bm{\mathrm{s}}}
\newcommand{\vv}{\bm{\mathrm{v}}}
\newcommand{\wv}{\bm{\mathrm{w}}}
\newcommand{\fp}{f^\prime}
\newcommand{\pp}{p^\prime}
\newcommand{\pvp}{\bm{\mathrm{p}}^\prime}
\newcommand{\rp}{r^\prime}
\newcommand{\rvp}{\bm{\mathrm{r}}^\prime}
\newcommand{\tp}{t^\prime}
\newcommand{\htil}{\tilde{h}}
\newcommand{\Deltat}{\tilde{\Delta}}
\newcommand{\kappat}{\tilde{\kappa}}
\newcommand{\psit}{\tilde{\psi}}
\newcommand{\varrhot}{\tilde{\varrho}}
\newcommand{\kappad}{\dot{\kappa}}
\newcommand{\kappatd}{\dot{\tilde{\kappa}}}
\newcommand{\nud}{\dot{\nu}}
\newcommand{\phid}{\dot{\phi}}
\newcommand{\varrhod}{\dot{\varrho}}
\newcommand{\rhod}{\dot{\rho}}
\newcommand{\varrhotd}{\dot{\tilde{\varrho}}}
\newcommand{\Deltah}{\hat{\Delta}}
\newcommand{\Fhat}{\hat{F}}
\newcommand{\jhat}{\hat{j}}
\newcommand{\jvh}{\hat{\bm{\mathrm{j}}}}
\newcommand{\nuh}{\hat{\nu}}
\newcommand{\phidh}{\hat{\dot{\phi}}}
\newcommand{\phih}{\hat{\phi}}
\newcommand{\rhoh}{\hat{\rho}}
\newcommand{\Vhat}{\hat{V}}
\newcommand{\iev}{\textit{ev}}
\newcommand{\iod}{\textit{od}}
\newcommand{\ire}{\textit{re}}
\newcommand{\iim}{\textit{im}}
\newcommand{\iext}{\textit{ext}}
\newcommand{\iloc}{\textit{loc}}
\begin{document}
\title{Dynamics of a trapped Fermi gas in the BCS phase}
\author{Michael Urban}
\affiliation{Institut de Physique Nucl{\'e}aire, F-91406 Orsay
C{\'e}dex, France}
\author{Peter Schuck}
\affiliation{Institut de Physique Nucl{\'e}aire, F-91406 Orsay
C{\'e}dex, France}
\begin{abstract}
We derive semiclassical transport equations for a trapped atomic Fermi
gas in the BCS phase at temperatures between zero and the superfluid
transition temperature. These equations interpolate between the two
well-known limiting cases of superfluid hydrodynamics at zero
temperature and the Vlasov equation at the critical one. The
linearized version of these equations, valid for small deviations from
equilibrium, is worked out and applied to two simple examples where
analytical solutions can be found: a sound wave in a uniform medium
and the quadrupole excitation in a spherical harmonic trap. In spite
of some simplifying approximations, the main qualitative results of
quantum mechanical calculations are reproduced, which are the
different frequencies of the quadrupole mode at zero and the critical
temperature and strong Landau damping at intermediate temperatures. In
addition we suggest a numerical method for solving the semiclassical
equations without further approximations.
\end{abstract}
\maketitle
%
\section{Introduction}
Due to improved cooling techniques, current experiments with trapped
fermionic atoms like $^6$Li or $^{40}$K reach very low temperatures of
the order of $T \approx 0.03 T_F$ \cite{Bartenstein}, where $T_F =
\epsilon_F/k_B$ denotes the degeneracy temperature. The main
motivation for these experiments is to study the so-called BEC-BCS
crossover by tuning the magnetic field around a Feshbach resonance,
thus changing the atom-atom scattering length $a$ from the repulsive
side ($a>0$) through the unitary limit ($a\to\infty$) to the
attractive side ($a<0$). On the BEC side, where the system forms a
Bose-Einstein condensate (BEC) of tightly bound molecules, as well as
on the BCS side of the crossover, where the atoms form Cooper pairs
which are very large compared with the mean distance between atoms,
one expects that the system is superfluid, provided the temperature
lies below a certain critical temperature $T_c$. Until now the
experiments have concentrated on the crossover region, but the low
temperatures which are currently reached suggest that future
experiments will also be able to study the superfluid BCS phase ($a <
0$ and $k_F |a| \ll 1$), although its critical temperature will be
extremely low.

In order to find signals for superfluidity, some recent experiments
with ultracold trapped Fermionic atoms looked at dynamical observables
like the expansion of the atom cloud after the trap has been switched
off \cite{OHara}, or collective oscillations of the cloud
\cite{Bartenstein,Kinast}. The theoretical interpretation of such
experiments is usually based on a theory called ``superfluid
hydrodynamics'' \cite{MenottiPedri,CozziniStringari,Stringari}, which
is valid for a superfluid at zero temperature if the trap potential is
sufficiently smooth to justify local-density approximation. Apart from
the latter condition, which is not necessarily fulfilled in the
experiments \cite{GrassoKhan}, it is also clear that experiments are
not done at zero temperature.

Very recently, Landau's two-fluid hydrodynamics has been used to
describe collective modes in trapped superfluid gases at non-zero
temperature \cite{TaylorGriffin}. In this approach, in the temperature
range $0<T<T_c$, a certain fraction of the atoms is not superfluid but
forms a normal-fluid component with density $\rho_n$, while the
remaining atoms with density $\rho_s = \rho - \rho_n$ still behave
like a superfluid. It is also assumed that the atoms undergo enough
collisions to be always in local equilibrium. This condition, however,
cannot be taken for granted. In \Refe{MassignanBruun} it was found that
even in the unitary limit, where the scattering length $a$ diverges,
the collision rate might be too low to ensure hydrodynamic behavior of
the normal phase. This is certainly true in the BCS phase, where $k_F
|a|$ and the temperature are so small that one can safely assume that
the system is in the so-called collisionless regime. Collisionless
means in this context that the collision rate $1/\tau$ is much smaller
than the trap frequency $\Omega$, i.e., an atom performs several
oscillations in the trap before colliding with another atom. Since the
frequencies of the collective oscillations are of the order of the
trap frequency $\Omega$, this implies that it is impossible to reach
local equilibrium during the oscillation.

Nevertheless the idea of a two-fluid model is useful in the
collisionless regime, too. It has been developed for this case in the
theory of superconductivity
\cite{Schrieffer,Leggett1,Leggett2,BetbederMatibet}. Similar approaches to
describe liquid $^3$He should be mentioned as well, although they are
more complicated because of the spin structure of the order parameter
\cite{SereneRainer,Woelfle}. Recently the two-fluid model has also
been applied to the case of trapped fermionic atoms in the BCS phase
\cite{UrbanSchuck,Nygaard}. Because of the possibility of
Fermi-surface deformations, the normal component of a collisionless
gas does not behave hydrodynamically, but more like an elastic body. A
semiclassical method for treating the Fermi surface deformation in a
normal Fermi gas is given by the Vlasov equation. The latter was used
with great success in nuclear physics, e.g., in order to describe
giant resonances in atomic nuclei \cite{RingSchuck}, and recently it
was also applied to trapped atomic Fermi gases in order to predict the
frequencies of collective modes in the collisionless regime
\cite{MenottiPedri}. Contrary to hydrodynamical equations, where all
quantities are local (i.e., functions of the spatial coordinate $\rv$
only), the Vlasov equation requires a phase-space description (i.e.,
the quantities are functions of $\rv$ and $\pv$). The aim of the
present article is to derive a hydrodynamical equation for the
superfluid component coupled to a Vlasov equation for the normal
component, interpolating between superfluid hydrodynamics at $T = 0$
and the usual Vlasov equation at $T = T_c$. In principle, as it was
done in the theory of liquid $^3$He \cite{SereneRainer}, one could
also think of including a collision term into this equation, in order
to treat systems which are neither collisionless nor hydrodynamical,
but somewhere in between. However, in the present article we will
restrict ourselves to the collisionless case.

Like the semiclassical description of the ground state (Thomas-Fermi
approximation), the semiclassical description of the dynamics of the
system can be expected to become more and more accurate if the number
of atoms in the trap increases. This was the main motivation for us to
develop the semiclassical approach presented here. A fully
quantum-mechanical description of the collective modes of a trapped
Fermi gas can be obtained, e.g., by the quasiparticle random-phase
approximation (QRPA), corresponding to the linearization of the
time-dependent Bogoliubov-de Gennes equations around equilibrium. The
latter are also known as time-dependent Hartree-Fock-Bogoliubov
(TDHFB) equations, especially in nuclear physics. QRPA calculations
become tremendously difficult and time-consuming if the number of
particles increases. At present, they are restricted to systems of
$\sim 10^4$ atoms \cite{GrassoKhan,BruunMottelson,OhashiGriffin},
while the numbers of atoms in the experiments are at least ten times
larger. In addition, all present QRPA calculations are done for the
case of spherically symmetric traps, while the traps used in the
experiments are generally not spherical. The numerical solution of the
QRPA equations without spherical symmetry seems to be almost
unfeasible, unless one reduces drastically the number of
particles. Therefore semiclassical approaches are at the moment the
only way to perform calculations for large numbers of atoms in
realistic trap geometries.

Our article is organized as follows. In \Sec{formalism} we will
present the formalism. Having derived a quasiparticle transport
equation in \Sec{transport}, an important point will be to work out
the linearized version of this equation in order to apply it to
oscillations around the equilibrium state. This is done in
\Sec{linearization}. In \Sec{limitingcases} we will show explicitly
that our equations indeed reproduce superfluid hydrodynamics and the
Vlasov equation in the limits of zero and critical temperature,
respectively. The next part, \Sec{examples}, is devoted to two simple
examples for which our equations can be solved more or less
analytically. The first example, discussed in \Sec{soundwave}, is a
sound wave in a uniform gas. The second one, described in
\Sec{quadrupole}, concerns a quadrupole oscillation of a harmonically
trapped gas with some additional simplifications. Finally, in
\Sec{summary} we will summarize and draw our conclusions.
\section{Formalism}
\label{formalism}
\subsection{Derivation of a quasiparticle transport equation}
\label{transport}
In this subsection we will derive a quasiparticle transport equation
for a superfluid gas of trapped fermionic atoms in the BCS
phase. Throughout this article we will assume that the two spin states
$\uparrow$ and $\downarrow$ are equally populated, which allows us to
remove the spin degree of freedom from the beginning. However, the
generalization to include the spin, which in fact would be necessary,
e.g., in order to describe spin waves or systems with unequal
populations, is straight-forward. In order to be in the BCS phase, the
atoms must have an attractive interaction, i.e., a negative scattering
length $a<0$, which on the other hand must be weak enough for the BCS
approximation to be valid.

Let us start by writing down the TDHFB equations \cite{RingSchuck}. To
that end we define the non-local normal and anomalous density
matrices,
\begin{gather}
\varrho(\rv,\rvp) 
 = \langle\psi^\dagger_\uparrow(\rvp) \psi_\uparrow(\rv)\rangle
 = \langle\psi^\dagger_\downarrow(\rvp) \psi_\downarrow(\rv)\rangle\,,\\
\kappa(\rv,\rvp) 
 = \langle\psi_\uparrow(\rvp) \psi_\downarrow(\rv)\rangle
 = -\langle\psi_\downarrow(\rvp) \psi_\uparrow(\rv)\rangle\,,
\end{gather}
where $\psi$ is the field operator. The single-particle hamiltonian
(minus the chemical potential $\mu$) reads
\begin{equation}
h = -\frac{\hbar^2\nablav^2}{2m}+V_\iext(\rv)+g\rho(\rv)-\mu\,,
\end{equation}
where $m$ is the atomic mass, $V_\iext(\rv)$ is the potential
of the trap, $g\rho(\rv)$ is the mean-field potential. The coupling
constant $g$ is related to the atom-atom scattering length $a$ by $g =
4\pi\hbar^2 a/m$ and the density per spin state $\rho(\rv)$ is just
equal to the local part of the density matrix,
\begin{equation}
\rho(\rv) = \varrho(\rv,\rv)\,.
\label{rhocoo}
\end{equation}
According to the usual regularization prescription \cite{BruunCastin},
the pairing gap is related to the anomalous density by
\begin{equation}
\Delta(\rv) = -g\lim_{s\to 0}\frac{d}{ds} s
  \,\kappa\Big(\rv+\frac{\sv}{2},\rv-\frac{\sv}{2}\Big)\,.
\label{gapeqcoo}
\end{equation}
Combining all quantities in the $2\times 2$ matrices
\begin{equation}
\mathcal{H} =
  \bigg(\begin{matrix}h &\Delta\\ \Delta^\dagger&
  -\bar{h}\end{matrix}\bigg)\,,
\qquad
\mathcal{R} = \bigg(\begin{matrix}\varrho &-\kappa\\
  -\kappa^\dagger& 1-\bar{\varrho}\end{matrix}\bigg)\,,
\end{equation}
where $\bar{\varrho}$ and $\bar{h}$ denote the time-reversed operators to
$\varrho$ and $h$, respectively, the TDHFB equation can be written in the
compact form \cite{RingSchuck}
\begin{equation}
i\hbar \dot{\mathcal{R}} = \big[\mathcal{H},\mathcal{R}\big]\,.
\label{tdhfb}
\end{equation}

In analogy to the derivation of the Vlasov equation in the normal
phase from the Hartree-Fock equation \cite{RingSchuck}, it is useful
to introduce the Wigner transform of the density matrix,
\begin{equation}
\varrho(\rv,\pv) =
  \int d^3 s e^{-i\pv\cdot\sv/\hbar}\,
  \varrho\big(\rv+\frac{\sv}{2},\rv-\frac{\sv}{2}\Big)\,.
\end{equation}
It is appealing, although strictly speaking not correct, to interprete
the function $\varrho(\rv,\pv)$ as a distribution function of particles
in phase space. In a completely analogous way we define the Wigner
transform of the anomalous density matrix, $\kappa(\rv,\pv)$, and the
Wigner transform of the hamiltonian, $h(\rv,\pv)$, which is equal to
the classical hamiltonian
\begin{equation}
h(\rv,\pv) = \frac{p^2}{2m}+V_\iext(\rv)+g\rho(\rv)-\mu\,.
\end{equation}
(For the sake of readability we are using the same symbol for the
operators and their Wigner transforms, but whenever there is a risk of
confusion we will write down the arguments.) \Eqs{rhocoo} and
(\ref{gapeqcoo}) can be written in terms of the Wigner transformed
quantities as follows:
\begin{gather}
\rho(\rv) = \int\frac{d^3p}{(2\pi\hbar)^3}\varrho(\rv,\pv)\,,
  \label{rhoeq}\\
\Delta(\rv) = -g\int \frac{d^3p}{(2\pi\hbar)^3}
  \Big(\kappa(\rv,\pv)-\frac{\Delta(\rv)}{p^2/m}\Big)\,.
\label{gapeq}
\end{gather}
We also need the Wigner transforms of the time-reversed operators
$\bar{\varrho}$ and $\bar{h}$, and the Wigner transforms of the adjoint
operators $\kappa^\dagger$ and $\Delta^\dagger$. To that end we recall
the general relations
\begin{equation}
\bar{A}(\rv,\pv) = A(\rv,-\pv)\,,\qquad 
[A^\dagger](\rv,\pv) = A^*(\rv,\pv)\,,
\end{equation}
which are valid for an arbitrary operator $A$. The usefulness of the
Wigner transform lies in the fact that, to first order in an expansion
into powers of $\hbar$, the Wigner transform of the product of two
operators $A$ and $B$ can be obtained according to
\begin{equation}
[AB](\rv,\pv) \approx A(\rv,\pv) B(\rv,\pv) + \frac{i\hbar}{2}
  \poisson{A(\rv,\pv)}{B(\rv,\pv)}\,,
\end{equation}
where $\poisson{\cdot}{\cdot}$ denotes te Poisson bracket
\begin{equation}
\poisson{A}{B} = \sum_{i=x,y,z}\Big(
  \frac{\partial A}{\partial r_i}\frac{\partial B}{\partial p_i}
  -\frac{\partial A}{\partial p_i}\frac{\partial B}{\partial r_i}\Big)\,.
\end{equation}
Applying this product rule to the Wigner transform of the TDHFB
equation (\ref{tdhfb}), one obtains four coupled equations:
\begin{widetext}
\begin{subequations}
\label{sctdhfb1}
\begin{gather}
i\hbar\varrhod = i\hbar\poisson{h}{\varrho}+2i\Im (\Delta^*\kappa)
  -i\hbar\Re\poisson{\Delta^*}{\kappa}\,,\\
i\hbar\kappad = (h+\bar{h})\kappa
  +\frac{i\hbar}{2}\poisson{h-\bar{h}}{\kappa}
  +\Delta(\varrho+\bar{\varrho}-1)
  -\frac{i\hbar}{2}\poisson{\Delta}{\varrho-\bar{\varrho}}\,,\\
i\hbar\kappad^* = -(h+\bar{h})\kappa^*
  +\frac{i\hbar}{2}\poisson{h-\bar{h}}{\kappa^*}
  -\Delta^*(\varrho+\bar{\varrho}-1)
  -\frac{i\hbar}{2}\poisson{\Delta^*}{\varrho-\bar{\varrho}}\,,\\
i\hbar\dot{\bar{\varrho}} = -i\hbar\poisson{\bar{h}}{\bar{\varrho}}
  +2i\Im (\Delta^*\kappa)
  +i\hbar\Re\poisson{\Delta^*}{\kappa}\,.
\end{gather}
\end{subequations}
\end{widetext}

In order to proceed further, it is useful to separate the collective
superfluid motion from the dynamics due to quasiparticle
excitations. This can be achieved by a gauge transformation,
$\psit(\rv) = \psi(\rv)\,\exp[i\phi(\rv)]$
\cite{BetbederMatibet,SereneRainer}. According to their definitions,
the normal and anomalous density matrices behave very differently
under this transformation. The corresponding Wigner transforms are
given by
\begin{gather}
\varrhot(\rv,\pv) =
  \varrho[\rv,\pv-\hbar\nablav\phi(\rv)]\,,\label{rhogauged}\\
\kappat(\rv,\pv) =
  \kappa(\rv,\pv)\, e^{2i\phi(\rv)}\,.\label{kappagauged}
\end{gather}
If the hamiltonian and the gap are changed according to
\begin{gather}
\htil(\rv,\pv) =
  \frac{[\pv-\hbar\nablav\phi(\rv)]^2}{2m}-\hbar\phid(\rv)
  +V(\rv)-\mu\,,\\
\Deltat(\rv) =
  \Delta(\rv)\,e^{2i\phi(\rv)}\,,
\end{gather}
the equation of motion of the gauge transformed quantities looks
exactly like \Eq{tdhfb}. The superfluid velocity is proportional to
the gradient of the phase of the gap. Hence, if we choose the gauge
transformation such that the transformed gap $\Deltat$ is real, we
have completely separated the collective motion of the superfluid
component from the motion due to quasiparticle excitations. A formal
argument for the necessity of this choice of the gauge is given in
\Refe{SereneRainer}.

From now on we will suppose that $\Deltat$ is real. Splitting $\varrhot$
and $\htil$ into time-even and time-odd parts,
\begin{gather}
\varrhot_\iev = \frac{1}{2}\big(\varrhot+\bar{\varrhot}\big)\,,\qquad
\varrhot_\iod = \frac{1}{2}\big(\varrhot-\bar{\varrhot}\big)\,,\\
\htil_\iev = \frac{1}{2}\big(\htil+\bar{\htil}\big)
  = \frac{p^2}{2m}
        +\frac{(\hbar\nablav\phi)^2}{2m}+V-\mu-\hbar\phid\,,
  \label{htilev}\\
\htil_\iod = \frac{1}{2}\big(\htil-\bar{\htil}\big)
  = -\frac{\hbar}{m}\pv\cdot\nablav\phi\,,
  \label{htilod}
\end{gather}
and $\kappat$ into real and imaginary parts,
\begin{gather}
\kappat_\ire = \Re\kappat\,,\qquad
\kappat_\iim = \Im\kappat\,,
\end{gather}
one can rewrite the gauge transformed version of the system of
equations (\ref{sctdhfb1}) as follows:
\begin{subequations}
\label{sctdhfb2}
\begin{gather}
\hbar\varrhotd_\iev = \hbar\poisson{\htil_\iev}{\varrhot_\iod}
  +\hbar\poisson{\htil_\iod}{\varrhot_\iev}+2\Deltat\kappat_\iim\,,
\label{sctdhfb2a}\\
\hbar\varrhotd_\iod = \hbar\poisson{\htil_\iev}{\varrhot_\iev}
  +\hbar\poisson{\htil_\iod}{\varrhot_\iod}
  -\hbar\poisson{\Deltat}{\kappat_\ire}\,,
\label{sctdhfb2b}\\
\hbar\kappatd_\ire = 2\htil_\iev\kappat_\iim
  +\hbar\poisson{\htil_\iod}{\kappat_\ire}
  -\hbar\poisson{\Deltat}{\varrhot_\iod}\,,
\label{sctdhfb2c}\\
\hbar\kappatd_\iim = -2\htil_\iev\kappat_\ire+\Deltat(1-2\varrhot_\iev)
  +\hbar\poisson{\htil_\iod}{\kappat_\iim}\,.
\label{sctdhfb2d}
\end{gather}
\end{subequations}

For a semiclassical $\hbar$ expansion it seems disturbing that these
equations mix different orders in $\hbar$. However, it is possible to
decouple the equations of motion for the leading-order quantities from
those of the higher-order ones. In order to show this, we expand
$\varrhot$ and $\kappat$ into powers of $\hbar$. Since the \Eqs{sctdhfb2}
themselves are only valid up to order $\hbar$, it does not make sense
to go beyond the first order in this series. From \Eqs{sctdhfb2a} and
(\ref{sctdhfb2c}) it is evident that $\kappat_\iim$ must be suppressed
by one power of $\hbar$ with respect to the other quantities. We
therefore write
\begin{gather}
\varrhot_{\iev,\iod} = \varrhot^{(0)}_{\iev,\iod}
  +\hbar\varrhot^{(1)}_{\iev,\iod}+\cdots\,,\\
\kappat_\ire = \kappat^{(0)}_\ire
  +\hbar\kappat^{(1)}_\ire+\cdots\,,\\
\kappat_\iim = \hbar\kappat^{(1)}_\iim+\cdots\,.
\end{gather}
Inserting these expansions into \Eqs{sctdhfb2a} -- (\ref{sctdhfb2d})
and retaining only the leading order in each equation [order $\hbar$
in the case of \Eqs{sctdhfb2a} -- (\ref{sctdhfb2c}), order $1$ in the
case of \Eq{sctdhfb2d}], one obtains
\begin{subequations}
\begin{gather}
\varrhotd^{(0)}_\iev =
  \poisson{\htil_\iev}{\varrhot^{(0)}_\iod}
  +\poisson{\htil_\iod}{\varrhot^{(0)}_\iev}
  +2\Deltat\kappat^{(1)}_\iim\,,
\label{sctdhfb3a}\\
\varrhotd^{(0)}_\iod = 
  \poisson{\htil_\iev}{\varrhot^{(0)}_\iev}
  +\poisson{\htil_\iod}{\varrhot^{(0)}_\iod}
  -\poisson{\Deltat}{\kappat^{(0)}_\ire}\,,
\label{sctdhfb3b}\\
\kappatd^{(0)}_\ire = 2\htil_\iev\kappat^{(1)}_\iim
  +\poisson{\htil_\iod}{\kappat^{(0)}_\ire}
  -\poisson{\Deltat}{\varrhot^{(0)}_\iod}\,,
\label{sctdhfb3c}\\
2\htil_\iev\kappat^{(0)}_\ire =
  \Deltat(1-2\varrhot^{(0)}_\iev)\,.
\label{sctdhfb3d}
\end{gather}
\end{subequations}
Only one of the higher-order quantities, namely
$\kappat^{(1)}_\iim$, appears in these equations, but it can
be expressed in terms of the leading-order quantities, e.g., with the
help of \Eq{sctdhfb3a}:
\begin{equation}
\kappat^{(1)}_\iim = 
  \frac{1}{2\Deltat}\big(\varrhotd^{(0)}_\iev
  -\poisson{\htil_\iev}{\varrhot^{(0)}_\iod}
  -\poisson{\htil_\iod}{\varrhot^{(0)}_\iev}\big)\,.
\label{kappatim}
\end{equation}
By taking a linear combination of \Eqs{sctdhfb3a} and
(\ref{sctdhfb3c}) one can eliminate $\kappat^{(1)}_\iim$. The
resulting equation reads
\begin{multline}
\htil_\iev\varrhotd^{(0)}_\iev
  -\Deltat\kappatd^{(0)}_\ire =
  E_\iev\poisson{E_\iev}{\varrhot^{(0)}_\iod}
  +\htil_\iev\poisson{\htil_\iod}{\varrhot^{(0)}_\iev}\\
  -\Deltat\poisson{\htil_\iod}{\kappat^{(0)}_\ire}\,,
\label{sctdhfb3e}
\end{multline}
where we have introduced the abbreviation
\begin{equation}
E_\iev = \sqrt{\htil^2_\iev+\Deltat^2}\,.
\end{equation}
\Eqs{sctdhfb3b}, (\ref{sctdhfb3d}), and (\ref{sctdhfb3e}) form a
system of three coupled equations for the three leading-order
quantities $\varrhot^{(0)}_\iev$, $\varrhot^{(0)}_\iod$,
and $\kappat^{(0)}_\ire$.

From now on we will suppress the index ``(0)'' and simply write
$\varrhot_\iev$, $\varrhot_\iod$, and $\kappat_\ire$ instead of
$\varrhot^{(0)}_\iev$, $\varrhot^{(0)}_\iod$, and
$\kappat^{(0)}_\ire$. The next step is to exploit
\Eq{sctdhfb3d} in order to reduce the number of unknown functions. To
that end we introduce a new phase-space function $\nu_\iev(\rv,\pv)$, the
so-called ``quasiparticle distribution function'', which is defined in
such a way that the two members of \Eq{sctdhfb3d} are equal to
$\htil_\iev \Deltat (1-2\nu_\iev)/E_\iev$. In other words, $\varrhot_\iev$ and
$\kappat_\ire$ can be expressed in terms of this function $\nu_\iev$ as
follows:
\begin{gather}
\varrhot_\iev = \frac{1}{2}-\frac{\htil_\iev}{2E_\iev}(1-2\nu_\iev)\,,
\label{rhotev}\\
\kappat_\ire = \frac{\Deltat}{2E_\iev}(1-2\nu_\iev)\,.
\label{kappatre}
\end{gather}
In fact, the definition of $\nu_\iev$ has been chosen such that these
relations resemble the well-known expressions for $\varrho$ and $\kappa$
in equilibrium, where $\nu_\iev$ has to be replaced by the Fermi
distribution function for quasiparticles, $f(E)$ (see
\Sec{linearization}). With the help of \Eqs{rhotev} and
(\ref{kappatre}) the remaining two equations, (\ref{sctdhfb3b}) and
(\ref{sctdhfb3e}), take the rather simple form
\begin{subequations}
\begin{gather}
\varrhotd_\iod = 
  \poisson{E_\iev}{\nu_\iev}+\poisson{\htil_\iod}{\varrhot_\iod}\,,
\label{sctdhfb4a}\\
\nud_\iev = 
  \poisson{E_\iev}{\varrhot_\iod}+\poisson{\htil_\iod}{\nu_\iev}\,.
\label{sctdhfb4b}
\end{gather}
\end{subequations}
Since the first of these equations is purely time-odd while the second
one is purely time-even, we can add both equations without any loss of
information. The result can be written as
\begin{equation}
\nud = \poisson{E}{\nu}\,,
\label{qpvlasov}
\end{equation}
where we have introduced the new functions
\begin{equation}
\nu = \nu_\iev+\varrhot_\iod\,,\qquad
E = E_\iev+\htil_\iod\,.
\end{equation}
\Eq{qpvlasov} resembles very much the usual Vlasov equation for the
normal Fermi gas, which can be written as $\varrhod =
\poisson{h}{\varrho}$. One just has to replace the distribution
function $\varrho$ by the quasiparticle distribution function $\nu$
and the hamiltonian $h$ by the quasiparticle energy $E$. It should be
mentioned that \Eq{qpvlasov} or similar kinetic equations have already
been derived in the literature several times. Probably for the first
time it was given by Betbeder-Matibet and Nozi\`eres
\cite{BetbederMatibet} in a linearized form for small deviations from
equilibrium. In order to be self-contained, we gave here our own way
to arrive at \Eq{qpvlasov}.

In order to obtain a closed system of equations, \Eq{qpvlasov} must be
complemented by an equation for the so-far unknown phase $\phi$. As
stated above, the phase is fixed by the requirement that the gauge
transformed gap $\Deltat$ is real, i.e., $\Im\Deltat = 0$. With the
help of the relation (\ref{kappatim}) and of the gap equation
(\ref{gapeq}), this can be rewritten as
\begin{equation}
\int \frac{d^3p}{(2\pi\hbar)^3}
  \big(\varrhotd_\iev-
  \poisson{\htil_\iev}{\varrhot_\iod}
  -\poisson{\htil_\iod}{\varrhot_\iev}\big) = 0\,.
\label{gapreal}
\end{equation}
As we will see in a moment, this is nothing but the continuity
equation. This observation confirms earlier statements in the
literature that the continuity equation should be used for the
determination of the phase \cite{BetbederMatibet}. In order to derive
the continuity equation from \Eq{gapreal}, we write down explicitly
the poisson brackets and integrate by parts. In this way we obtain
\begin{equation}
\int \frac{d^3p}{(2\pi\hbar)^3} \Big(\varrhotd
  +\nablav\cdot\varrhot\frac{\pv-\hbar\nablav\phi}{m}\Big) 
  = 0\,.
\label{contigauged}
\end{equation}
Using \Eq{rhogauged} and changing the integration variable according
to $\pv\to\pv+\hbar\nablav\phi$, this can be transformed into the
usual continuity equation,
\begin{equation}
\rhod(\rv) + \nablav\cdot\jv(\rv) = 0\,,
\end{equation}
with
\begin{equation}
\jv(\rv) = \int \frac{d^3p}{(2\pi\hbar)^3} \frac{\pv}{m}\varrho(\rv,\pv)\,.
\end{equation}
\subsection{Linearization around equilibrium}
\label{linearization}
From now on we will assume that the external potential
$V_\iext$ can be written as
\begin{equation}
V_\iext = V_{0\iext} + V_{1\iext}\,,
\end{equation}
where $V_{0\iext}$ is time-independent and $V_{1\iext}$
can be considered as a small perturbation. The equilibrium quantities
corresponding to the static potential $V_{0\iext}$ will be
marked by an index ``0'', e.g.,
\begin{gather}
\nu_0(\rv,\pv) = f[E_0(\rv,\pv)]\,,\\
\phi_0(\rv) = 0\,,
\end{gather}
where $f(E)$ denotes the Fermi function
\begin{equation}
f(E) = \frac{1}{e^{E/(k_B T)}+1}
\end{equation}
and
\begin{gather}
E_0(\rv,\pv) = \sqrt{h_0^2(\rv,\pv)+\Delta_0^2(\rv)}\,,\\
h_0(\rv,\pv) = \frac{p^2}{2m}+V_{0\iext}(\rv)+g\rho_0(\rv)-\mu\,,
\end{gather}
etc. Our aim is to calculate the small deviations from equilibrium
induced by the perturbation $V_{1\iext}$, which will be marked
by an index ``1''. To that end we linearize the equation of motion
(\ref{qpvlasov}) for the quasiparticle distribution function:
\begin{equation}
\nud_1 - \poisson{E_0}{\nu_1} 
  = \fp(E_0)\poisson{E_1}{E_0}\,,
\label{qpvlasovlin1}
\end{equation}
where $\fp(E_0) = df/dE_0$. We also linearize the continuity
equation (\ref{contigauged}):
\begin{equation}
\rhod_1(\rv) + \nablav\cdot\jv_{1\nu}(\rv)-\frac{\hbar}{m}
  \nablav\cdot\rho_0(\rv)\nablav\phi_1(\rv) = 0\,,
\end{equation}
with
\begin{equation}
\jv_{1\nu}(\rv) = \int \frac{d^3p}{(2\pi\hbar)^3} \frac{\pv}{m}
  \nu_1(\rv,\pv)\,.
\label{j1nudef}
\end{equation}

In order to have a closed system of equations, we must express
$E_1(\rv,\pv)$ and $\rho_1(\rv)$ in terms of equilibrium quantities,
the perturbation $V_{1\iext}(\rv)$, and the unknown functions
$\nu_1(\rv,\pv)$ and $\phi_1(\rv,\pv)$. Linearizing $E(\rv,\pv)$, one
obtains
\begin{equation}
E_1 = \frac{h_0}{E_0}\htil_{1\iev}+\frac{\Delta_0}{E_0}\Deltat_1
  +\htil_{1\iod}\,,
\end{equation}
with
\begin{gather}
\htil_{1\iev}(\rv,\pv) = V_{1\iext}(\rv)+g\rho_1(\rv)-
  \hbar\phid_1(\rv)\,,\\
\htil_{1\iod}(\rv,\pv) = -\frac{\hbar}{m}\pv\cdot\nablav\phi_1(\rv)\,.
\end{gather}

The most difficult part is to derive the expressions for $\rho_1(\rv)$
and $\Deltat_1(\rv)$. We start by linearizing \Eqs{rhotev} and
(\ref{kappatre}):
\begin{widetext}
\begin{gather}
\varrhot_{1\iev} = \frac{h_0}{E_0}\nu_{1\iev}
  +\frac{1-2f(E_0)}{2E_0^3}[-\Delta_0^2(V_{1\iext}+g\rho_1-
  \hbar\phid_1)+h_0\Delta_0\Deltat_1]\,,
\label{rhot1ev}\\
\kappat_{1\ire} = -\frac{\Delta_0}{E_0}\nu_{1\iev}
  -\frac{1-2f(E_0)}{2E_0^3}[h_0\Delta_0(V_{1\iext}+g\rho_1-
  \hbar\phid_1)
  +\Delta_0^2\Deltat_1]+\frac{1-2f(E_0)}{2E_0}\Deltat_1\,.
\label{kappat1re}
\end{gather}
According to \Eqs{rhoeq} and (\ref{gapeq}), $\rho_1$ and $\Deltat_1$
can be obtained by integrating \Eqs{rhot1ev} and (\ref{kappat1re})
over $\pv$. This gives a coupled system of two linear equations,
\begin{subequations}
\begin{gather}
\rho_1(\rv) = \rho_{1\nu}(\rv)
    -A(\rv)[V_{1\iext}(\rv)+g\rho_1(\rv)-\hbar\phid_1(\rv)]
    +B(\rv)\Deltat_1(\rv)\,,
  \label{rho1a}\\
\Deltat_1(\rv) = \Delta_{1\nu}(\rv)
    +gB(\rv) [V_{1\iext}(\rv)+g\rho_1(\rv)-\hbar\phid_1(\rv)]
    +[gA(\rv)+1] \Deltat_1(\rv)\,,
\label{deltat1a}
\end{gather}
\end{subequations}
\end{widetext}
where the gap equation (\ref{gapeq}) for the equilibrium case has been
used in the derivation of the last term, and the following
abbreviations have been introduced:
\begin{gather}
\rho_{1\nu}(\rv) = \int\frac{d^3p}{(2\pi\hbar)^3}
  \frac{h_0(\rv,\pv)}{E_0(\rv,\pv)}\nu_{1\iev}(\rv,\pv)\,,
  \label{rho1nudef}\\
\Delta_{1\nu}(\rv) = g\int\frac{d^3p}{(2\pi\hbar)^3}
  \frac{\Delta_0(\rv)}{E_0(\rv,\pv)}\nu_{1\iev}(\rv,\pv)\,,
  \label{delta1nudef}\\
A(\rv) = \Delta_0^2(\rv)\int\frac{d^3p}{(2\pi\hbar)^3}
  \frac{1-2f[E_0(\rv,\pv)]}{2E_0^3(\rv,\pv)}\,,
  \label{intega}\\
B(\rv) = \Delta_0(\rv)\int\frac{d^3p}{(2\pi\hbar)^3}h_0(\rv,\pv)
    \frac{1-2f[E_0(\rv,\pv)]}{2E_0^3(\rv,\pv)}\,.
  \label{integb}
\end{gather}

Below we will show that the coefficient $B$ is negligible compared
with the coefficient $A$. In the limit $B\to 0$ the two equations
(\ref{rho1a}) and (\ref{deltat1a}) are decoupled and can immediately
be solved for $\rho_1$ and $\Deltat_1$:
\begin{gather}
\rho_1(\rv) = \frac{\rho_{1\nu}(\rv)-A(\rv) [V_{1\iext}(\rv)
  -\hbar\phid_1(\rv)]}{1+gA(\rv)}\,,
  \label{rho1}\\
\Deltat_1(\rv) = -\frac{\Delta_{1\nu}(\rv)}{g A(\rv)}\,.
  \label{deltat1}
\end{gather}
In addition, we point out that $\Deltat_1=0$ if $\Delta_0=0$, which is
not evident from \Eq{deltat1} but can be derived from \Eq{kappat1re}.

\begin{figure}[t]
\includegraphics[width=6.5cm]{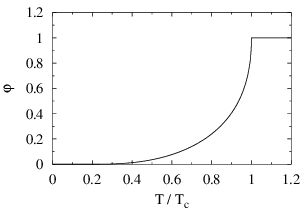}
\caption{Temperature dependence of the function $\varphi$ defined in
\Eq{defphi}.\label{figphi}}
\end{figure}

We will now calculate the coefficients $A$ and $B$ for the case that
both $\Delta_0(\rv)$ and $k_BT$ are small compared with the local Fermi
energy
\footnote{Note that in the case of a trapped system, the condition
$\Delta_0(\rv) \ll \epsilon_F(\rv)$ is automatically fulfilled
everywhere in the trap if it is valid at the center.}
\begin{equation}
\epsilon_F(\rv) = \frac{p_F^2(\rv)}{2m}
  =\mu-V_{0\iext}(\rv)-g\rho_0(\rv)\,.
\end{equation}
In this case, the relevant contributions to the integrals
(\ref{intega}) and (\ref{integb}) come from momenta near the Fermi
surface. As usual, the integrals over $p$ can be simplified by
transforming them into integrals over the energy variable $\xi =
p^2/2m-\epsilon_F(\rv)$ and approximating the density of states by its
value at the Fermi energy, i.e., $p^2 dp\approx mp_F(\rv) d\xi$. For
the coefficient $A$, one obtains in this way
\begin{equation}
A(\rv) = \frac{mp_F(\rv)}{2\pi^2\hbar^3}[1-\varphi(\rv)]\,,
\end{equation}
where the function $\varphi$ describes the temperature dependence:
\begin{equation}
\varphi(\rv) = -\int d\xi\frac{\xi^2}{E^2}\fp(E)
  \Big|_{E=\sqrt{\xi^2+\Delta_0^2(\rv)}}\,.
\label{defphi}
\end{equation}
One can show that $\varphi = 0$ for $T = 0$ and $\varphi = 1$ for
$\Delta_0 = 0$. In all other cases, the function $\varphi$ must be
evaluated numerically. From its definition one can see that $\varphi$
depends on $\rv$ only through the dimensionless parameter
$T/T_c(\rv)$, where $T_c(\rv) = 0.57 \Delta_0(\rv;T=0)/k_B$ is the
local critical temperature
\footnote{It is evident that $\varphi$ is a function of
$\Delta_0(\rv)/(k_BT)$, but $\Delta_0(\rv)$ can in turn be written as
$k_B T_c(\rv)$ times a universal function of $T/T_c(\rv)$.}.
For illustration, the numerical result for $\varphi$ as a function of
this parameter is shown in \Fig{figphi}.

If one applies the same method to the coefficient $B$, one obtains
$B=0$. This is because the integrand in \Eq{integb} is odd in $\xi$ if
one neglects the energy dependence of the density of states. Already
from this argument one can conclude that the coefficient $B$ must be
suppressed by at least one power of $\Delta_0/\epsilon_F$ or
$T/\epsilon_F$. Indeed, after a rather lengthy and delicate analysis
one finds
\begin{equation}
B(\rv) = \frac{\Delta_0(\rv)}{2\epsilon_F(\rv)}\Big(
  \frac{mp_F(\rv)}{2\pi^2\hbar^3}[2+\varphi(\rv)]-\frac{1}{g}\Big)\,.
\end{equation}
This is the justification for neglecting the coefficient $B$ when
solving \Eqs{rho1a} and (\ref{deltat1a}).

Finally, let us put everything together and give a concise summary of
the system of equations which has to be solved. First of all, there is
the equation of motion (\ref{qpvlasovlin1}) for the quasiparticle
distribution function. After the Poisson bracket on the r.h.s. has
been written down explicitly, it can be transformed into
\begin{widetext}
\begin{multline}
\nud_1-\poisson{E_0}{\nu_1} = 
  -\frac{\fp(E_0)}{m}\Big[-\pv\cdot\nablav
    \frac{V_{1\iext}+g\rho_{1\nu}-\hbar\phid_1}{1+gA}
  +\frac{\Delta_0}{E_0^2}\pv\cdot\nablav
    \frac{\Delta_0(V_{1\iext}+g\rho_{1\nu}-\hbar\phid_1)}{1+gA}
  +\frac{h_0}{E_0^2}\pv\cdot\nablav
    \frac{\Delta_0\Delta_{1\nu}}{gA}\\
  +\frac{\hbar}{m}\frac{h_0}{E_0}(\pv\cdot\nablav)^2\phi_1
  -\hbar\Big(\frac{h_0}{E_0}\nablav (V_{0\iext}+g\rho_0)+
    \frac{\Delta_0}{E_0}\nablav\Delta_0\Big)\cdot\nablav\phi_1\Big]\,.
\label{qpvlasovlin}
\end{multline}
The second equation is the continuity equation
\begin{equation}
\frac{\rhod_{1\nu}(\rv)
    -A(\rv)[\dot{V}_{1\iext}(\rv)-\hbar\ddot{\phi}_1(\rv)]}{1+gA(\rv)}
  +\nablav\cdot\jv_{1\nu}(\rv)
  -\frac{\hbar}{m}\nablav\cdot\rho_0(\rv)\nablav\phi_1(\rv) = 0\,.
\label{contilin}
\end{equation}
\end{widetext}
The definitions of $\rho_{1\nu}$, $\Delta_{1\nu}$, and $\jv_{1\nu}$ in
terms of $\nu_1$ are given by \Eqs{rho1nudef}, (\ref{delta1nudef}),
and (\ref{j1nudef}).
\subsection{Limiting cases}
\label{limitingcases}
We are now going to check that our equations reproduce superfluid
hydrodynamics and the Vlasov equation in the cases $T = 0$ and $T\geq
T_c$, respectively. In the limit of zero temperature, \Eq{qpvlasovlin}
becomes extremely simple since $f(E)=0$ and therefore the
r.h.s. of \Eq{qpvlasovlin} vanishes identically. The corresponding
solution is of course $\nu_1=0$
\footnote{This trivial solution is unique if we assume that the system
was in equilibrium at the moment when the time-dependent perturbation
was switched on.},
which implies $\rho_{1\nu}=\Delta_{1\nu}=\jv_{1\nu}=0$. As a
consequence, the continuity equation (\ref{contilin}) reduces to
\begin{equation}
\cfrac{\dot{V}_{1\iext}(\rv)-\hbar\ddot{\phi}_1(\rv)}
    {\cfrac{2\pi^2\hbar^3}{mp_F(\rv)}+g}
  +\frac{\hbar}{m}\nablav\cdot\rho_0(\rv)\nablav\phi_1(\rv) = 0\,.
\label{hydro}
\end{equation}
Here we have used the explicit expression for $A(\rv)$ and the fact
that $\varphi=0$ at zero temperature.

How does \Eq{hydro} compare to superfluid hydrodynamics? The
continuity and Euler equations of superfluid hydrodynamics can be
written as \cite{CozziniStringari}:
\begin{gather}
\rhod(\rv)+\nablav\cdot\rho(\rv)\vv(\rv)=0\,,
\label{hydroconti1}\\
\dot{\vv}(\rv) = -\nablav\Big(\frac{\vv^2(\rv)}{2}
  +\frac{V_\iext(\rv)}{m}+\frac{\mu_\iloc(\rv)}{m}\Big)\,,
\label{hydroeuler1}
\end{gather}
where $\vv(\rv)$ denotes the velocity field and $\mu_\iloc(\rv)$ is
the local chemical potential, which in the BCS phase
($\Delta\ll\epsilon_F$) is related to the density $\rho(\rv)$ by the
Thomas-Fermi relation
\begin{equation}
\mu_\iloc(\rv) = 
  \frac{p_F^2(\rv)}{2m}+g\rho(\rv)\,,
\end{equation}
with
\begin{equation}
p_F(\rv) = \hbar[6\pi^2\rho(\rv)]^{1/3}\,.
\end{equation}
Writing the irrotational velocity field in the form
\begin{equation}
\vv(\rv) = -\frac{\hbar}{m}\nablav\phi(\rv)
\end{equation}
and linearizing \Eqs{hydroconti1} and (\ref{hydroeuler1}) around
equilibrium, one obtains
\begin{gather}
\rhod_1(\rv)-\frac{\hbar}{m}\nablav\cdot\rho_0(\rv)\nablav\phi_1(\rv)=0\,,
  \label{hydroconti}\\
\hbar\phid(\rv)=V_{1\iext}(\rv)
  +\Big(\frac{2\pi^2\hbar^3}{mp_F(\rv)}+g\Big)\rho_1(\rv)\,.
  \label{hydroeuler}
\end{gather}
Solving \Eq{hydroeuler} for $\rho_1$ and inserting the result into
\Eq{hydroconti}, one reproduces exactly \Eq{hydro}. This can be seen
as an alternative to the recent derivation of superfluid hydrodynamics
from the underlying microscopic theory in \Refe{ToniniCastin}.

The analysis of the other limit, $T\geq T_c$, is more difficult. In
this limit, the gap $\Delta_0$ vanishes and consequently
\begin{gather}
E_0(\rv,\pv) = |h_0(\rv,\pv)|\,,\\
\nu_{1\iev}(\rv,\pv) = 
  \sgn [p-p_F(\rv)] \varrhot_{1\iev}(\rv,\pv)\,.
\end{gather}
In addition, one has $\varphi(\rv) = 1$, $A(\rv) = 0$, $\rho_1(\rv) =
\rho_{1\nu}(\rv)$, and $\Deltat_1(\rv) = \Delta_{1\nu}(\rv) =
0$. Using these relations, and considering separately the two cases
$p<p_F$ (i.e., $h_0<0$) and $p>p_F$ (i.e., $h_0>0$), one can convince
oneself that \Eqs{qpvlasovlin} and (\ref{contilin}) reduce to
\begin{multline}
\varrhotd_1-\poisson{h_0}{\varrhot_1} = \frac{\fp(h_0)}{m}\Big(
  -\pv\cdot\nablav(V_{1\iext}+g\rho_1-\hbar\phid_1)\\
  +\frac{\hbar}{m}(\pv\cdot\nablav)^2\phi_1 
  -\hbar[\nablav(V_{0\iext}+g\rho_0)]\cdot\nablav\phi_1\Big)\,.
\label{qpvlasovlintc}
\end{multline}
and
\begin{equation}
\rhod_1(\rv)+\nablav\cdot\jv_{1\nu}(\rv)-
  \frac{\hbar}{m}\nablav\cdot\rho_0(\rv)\nablav\phi_1(\rv)=0\,.
\label{contilintc}
\end{equation} 
As we will see in a moment, these two equations are not independent of
each other. Hence, they do not allow to determine $\varrhot_1(\rv,\pv)$
and $\phi_1(\rv)$ in a unique way. This is in fact very reasonable
since the condition $\Im\Deltat = 0$ fixing the phase $\phi$ becomes
meaningless above $T_c$, where $\Deltat = 0$, and therefore the
function $\phi$ should be completely arbitrary in this case. The
relevant physical quantity, which of course should be unique, is
$\varrho_1(\rv,\pv)$. Linearizing \Eq{rhogauged} and using
$\varrho_0(\rv,\pv) = f[h_0(\rv,\pv)]$, we can express
$\varrhot_1(\rv,\pv)$ in terms of $\varrho_1(\rv,\pv)$ as follows:
\begin{equation}
\varrhot_1(\rv,\pv) = \varrho_1(\rv,\pv)-\frac{\hbar}{m}\fp[h_0(\rv,\pv)]
  \pv\cdot\nablav\phi_1\,.
\label{rhogaugedlin}
\end{equation}
If we insert this into \Eq{qpvlasovlintc}, all terms containing the phase
$\phi_1$ drop out, and we are left with
\begin{equation}
\varrhod_1-\poisson{h_0}{\varrho_1} =
  \fp(h_0)\frac{\pv}{m}\cdot\nablav (V_{1\iext}+g\rho_1)\,.
\label{vlasovlin}
\end{equation}
This is nothing but the linearized form of the Vlasov equation,
\begin{equation}
\varrhod_1-\poisson{h_0}{\varrho_1} = \poisson{h_1}{\varrho_0}\,,
\end{equation}
with $h_1 = V_{1\iext}+g\rho_1$. It remains to check that the
continuity equation (\ref{contilintc}) is satisfied for arbitrary
functions $\phi_1$, if $\varrho_1$ fulfills \Eq{vlasovlin}. To that end,
we multiply \Eq{vlasovlin} by $\pv$ and integrate over $\pv$, which
leads to the usual continuity equation
\begin{equation}
\rhod_1(\rv)+\nablav\cdot\jv_1(\rv) = 0\,.
\label{normalcontilin}
\end{equation}
With the help of \Eq{rhogaugedlin} the current $\jv_1$ can be written
as
\begin{equation}
\jv_1(\rv) = \jv_{1\nu}(\rv)-\frac{\hbar}{m}\rho_0(\rv)\nablav\phi_1(\rv)\,.
\label{jv1decomposition}
\end{equation}
Inserting this into \Eq{normalcontilin}, we indeed recover
\Eq{contilintc}. Since we did not make any assumptions about the
function $\phi_1(\rv)$, we conclude that it is completely arbitrary, as
it should be.
\section{Simple examples}
\label{examples}
\subsection{Sound wave in a uniform system}
\label{soundwave}
In this subsection we are considering a particularly simple
excitation, namely a sound wave traveling through a uniform
medium. This case has already been studied by Leggett \cite{Leggett2}
many years ago (except for the numerical evaluation of the integrals)
by using the standard techniques of normal and anomalous Green's
functions. The purpose of the present subsection is therefore to check
that our apparently very complicated equations (\ref{qpvlasovlin}) and
(\ref{contilin}) correctly interpolate between the limits of zero and
critical temperature.

Since the medium is assumed to be uniform, the equilibrium quantities
do not depend on $\rv$. We consider an excitation operator of the form
\begin{equation}
V_{1\iext}(\rv;t)=\Vhat_{1\iext}\,
e^{i\kv\cdot\rv-i\omega t}\,.
\end{equation}
As usual, in order to ensure that the perturbation vanishes for $t\to
-\infty$, one can assume that $\omega$ has an infinitesimal positive
imaginary part. From translational invariance it is clear that all
quantities describing the deviations from equilibrium will also have
the form of a plane wave, with the same wave vector $\kv$ and
frequency $\omega$ as the excitation. Like $\Vhat_{1\iext}$, the
amplitudes will be marked by a hat over the corresponding
symbol. Concerning the phase $\phi_1$, it turns out to be convenient
to parametrize it in the form
\begin{equation}
\phi_1(\rv;t)=\phidh_1\,\frac{i}{\omega}\,
  e^{i\kv\cdot\rv-i\omega t}\,.
\end{equation}
The Poisson bracket on the l.h.s. of \Eq{qpvlasovlin} now becomes
\begin{equation}
\poisson{E_0}{\nu_1} = -i\frac{h_0}{E_0}\frac{\pv\cdot\kv}{m}\nuh_1
  e^{i\kv\cdot\rv-i\omega t}\,,
\end{equation}
and \Eq{qpvlasovlin} can easily be solved for $\nuh_1$:
\begin{widetext}
\begin{equation}
\nuh_1 = 
  \cfrac{-\fp(E_0)\cfrac{\pv\cdot\kv}{m\omega}\,\cfrac{h_0}{E_0}\Big(
      \cfrac{h_0}{E_0}\,\cfrac{\Vhat_{1\iext}-
        \hbar\phidh_1+g\rhoh_{1\nu}}{1+gA}
      -\cfrac{\Delta_0}{E_0}\,\cfrac{\Deltah_{1\nu}}{gA}
      +\cfrac{\pv\cdot\kv}{m\omega}\,\hbar\phidh_1\Big)}
   {1-\cfrac{h_0}{E_0}\,\cfrac{\pv\cdot\kv}{m\omega}}\,.
\label{nuhatsw}
\end{equation}
\end{widetext}

Of course, the quantities $\rhoh_{1\nu}$ and $\Deltah_{1\nu}$ on the
r.h.s. depend themselves on $\nuh_1$. Therefore the next step consists
in inserting this expression for $\nuh_1$ into \Eqs{rho1nudef} and
(\ref{delta1nudef}). The integrals over the angle between $\pv$ and
$\kv$ can be evaluated in closed form. For the remaining integrals
over $p$, we will again exploit the fact that the gap and the
temperature are much smaller than the Fermi energy, as we did already
in \Sec{linearization}. We thus replace $p^2dp$ by $mp_Fd\xi$, and in
the integrand we replace $p$ by $p_F$, except for $h_0$ and $E_0$,
which must be replaced by $\xi$ and $\sqrt{\xi^2+\Delta_0^2}$,
respectively. Like the coefficient $B$ in \Sec{linearization}, the
integrals which lead to the coupling between the equations for
$\rhoh_{1\nu}$ and $\Deltah_{1\nu}$ are zero within this
approximation, i.e., they are of higher order in $\Delta/\epsilon_F$
or $T/\epsilon_F$ and can be neglected. The resulting equation for
$\rhoh_{1\nu}$ reads
\begin{equation}
\rhoh_{1\nu} = -\frac{mp_F}{2\pi^2\hbar^3}\Big(
 \frac{\Vhat_{1\iext}-\hbar\phidh_1+g\rhoh_{1\nu}}{1+gA}\,I_2(s)
  +\hbar\phidh_1 I_0(s)\Big)\,.
\label{rho1nusw}
\end{equation}
Here we have introduced the abbreviation
\begin{equation}
I_n(s) = -\int d\xi \fp(E)
  \Big(\frac{\xi}{E}\Big)^n \Big[1-\frac{sE}{\xi}
  \arctanh\Big(\frac{\xi}{sE}\Big)\Big]\,,
\label{indef}
\end{equation}
where $E = \sqrt{\xi^2+\Delta_0^2}$, and $s$ denotes the dimensionless
ratio of the sound velocity $c = \omega/k$ and the Fermi velocity $v_F
= p_F/m$,
\begin{equation}
s = \frac{c}{v_F} = \frac{m\omega}{p_F k}\,.
\end{equation}
Although not marked explicitly, $I_n(s)$ depends not only on $s$ but
also on the ratio $T/T_c$ (analogously to the function
$\varphi$). Note that the integrals $I_n(s)$ have a branch cut along
the real axis from $s = -1$ to $s=1$. The infinitesimal imaginary part
of $\omega$, i.e., of $s$, fixes the sign of the imaginary part of
$I_n(s)$.

\begin{figure}[t]
\includegraphics[width=6.5cm]{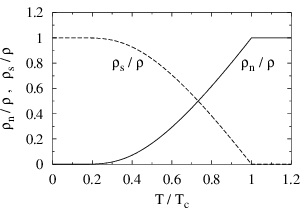}
\caption{Temperature dependence of $\rho_n/\rho_0$ (solid line) and
$\rho_s/\rho_0$ (dashed line).\label{figrhons}}
\end{figure}

Until now we have one equation for two unknown quantities,
$\rhoh_{1\nu}$ and $\phidh_1$. The second equation can be obtained
from the continuity equation (\ref{contilin}). It is evident that the
current $\jv_{1\nu}$ flows in longitudinal direction, such that it can
be written in the form
\begin{equation}
\jv_{1\nu}(\rv;t) = \jhat_{1\nu}\,\frac{\kv}{k}\,e^{i\kv\cdot\rv-i\omega t}\,.
\end{equation}
We will now express $\jhat_{1\nu}$ in terms of $\Vhat_{1\iext}$,
$\phidh_1$, and $\rhoh_{1\nu}$ by inserting \Eq{nuhatsw} into
\Eq{j1nudef}. The integration over $\pv$ is done as explained above for
the case of $\rhoh_{1\nu}$, and the result reads
\begin{multline}
\jhat_{1\nu}=-\frac{mcp_F}{2\pi^2\hbar^3}\Big(
  \frac{\Vhat_{1\iext}-\hbar\phidh+g\rhoh_{1\nu}}{1+gA}\,I_0(s)\\
    +\hbar\phidh_1 I_{-2}(s)\Big)
  -\frac{\rho_n\hbar\phidh_1}{mc}\,.
\label{jsw}
\end{multline}
In the last term, we have introduced the ``normal density'' of the
system, $\rho_n$, which is given by
\begin{equation}
\rho_n = \rho_0-\rho_s = -\rho_0\int d\xi \fp(E_0)\,.
\end{equation}
Correspondingly, $\rho_s$ is the ``superfluid density''.  Note that
the ratios $\rho_n/\rho_0$ and $\rho_s/\rho_0$ depend only on one
parameter, namely $T/T_c$. The numerical results for $\rho_n/\rho_0$
and $\rho_s/\rho_0$ as functions of $T/T_c$ are shown in
\Fig{figrhons}.

Inserting \Eq{jsw} into the continuity equation
(\ref{contilin}), one obtains the second equation which is needed for
determining $\rhoh_{1\nu}$ and $\phidh_1$:
\begin{widetext}
\begin{equation}
\frac{\rhoh_{1\nu}-A(\Vhat_{1\iext}-\hbar\phidh_1)}{1+gA}
  +\frac{mp_F}{2\pi^2\hbar^3}\Big(
  \frac{\Vhat_{1\iext}-\hbar\phidh+g\rhoh_{1\nu}}{1+gA}\,I_0(s)
    +\hbar\phidh_1 I_{-2}(s)\Big)-\frac{\rho_s\hbar\phidh_1}{mc^2}=0\,.
\label{contisw}
\end{equation}
In principle we could now solve \Eqs{rho1nusw} and (\ref{contisw}) for
the two unknown variables $\rhoh_{1\nu}$ and $\phidh_1$. However, it
is more transparent to use the amplitude of the total density
oscillations, $\rhoh_1$, as variable instead of the auxiliary quantity
$\rho_{1\nu}$. Expressing $\rhoh_{1\nu}$ in terms of $\rhoh_1$ with
the help of \Eq{rho1}, we rewrite \Eqs{rho1nusw} and (\ref{contisw})
as
\begin{subequations}
\label{rho1phi1sw}
\begin{gather}
\Big(1+\frac{gmp_F}{2\pi^2\hbar^3}[1-\varphi+I_2(s)]\Big)\rhoh_1
  -\frac{mp_F}{2\pi^2\hbar^3}[1-\varphi+I_2(s)-I_0(s)]\hbar\phidh_1
  = -\frac{mp_F}{2\pi^2\hbar^3}[1-\varphi+I_2(s)]\Vhat_{1\iext}\,,\\
\Big(1+\frac{gmp_F}{2\pi^2\hbar^3}\,I_0(s)\Big)\rhoh_1
  +\frac{mp_F}{2\pi^2\hbar^3}\Big(I_{-2}(s)-I_0(s)
    -\frac{1}{3s^2}\,\frac{\rho_s}{\rho_0}\Big)\hbar\phidh_1
  =-\frac{mp_F}{2\pi^2\hbar^3}\,I_0(s)\Vhat_{1\iext}\,.
\end{gather}
\end{subequations}
\end{widetext}
It is straight-forward to solve this $2\times 2$ system of equations
for $\rhoh_1$. Let us introduce the response function $\Pi$, defined
such that
\begin{equation}
\rhoh_1 = \frac{mp_F}{2\pi^2\hbar^3} \Pi(s;T/T_c;k_F a) \Vhat_1\,.
\end{equation}
The first term has been factored out in order to make $\Pi$
dimensionless. From the system of equations (\ref{rho1phi1sw}) one can
see that $\Pi$ is a function of $s$ and the two parameters $T/T_c$ and
\begin{equation}
k_Fa = \frac{gmp_F}{4\pi\hbar^3}\,.
\end{equation}
The explicit expression for $\Pi$ can most conveniently been expressed
in the form
\begin{equation}
\Pi(s;T/T_c;k_F a) = \cfrac{\Pi_0(s;T/T_c)}
  {1-\cfrac{2k_F a}{\pi}\,\Pi_0(s;T/T_c)}\,,
\end{equation}
where $\Pi_0$ is the response function in the limit $k_F a \to 0$:
\begin{equation}
\Pi_0 = \cfrac{(1-\varphi+I_2)
    \Big(\cfrac{1}{3s^2}\,\cfrac{\rho_s}{\rho_0}-I_{-2}\Big)+I_0^2}
  {1-\cfrac{1}{3s^2}\,\cfrac{\rho_s}{\rho_0}-\varphi+I_2+I_{-2}-2I_0}\,.
\end{equation}

Note that these expressions coincide exactly with the quantum
mechanical result in the long-wavelength and low-frequency limit as
given by Eqs.\@ (68) and (69) of \Refe{Leggett2}. In order to see this,
it is sufficient to observe that after integration over the solid
angle the quantities $\alpha$, $\zeta$, and $\eta$ defined in Eq.\@
(65) of \Refe{Leggett2} can be expressed in terms of our integrals as
$\alpha = (1 - \varphi+I_2-I_0) / 2$, $\zeta = [\rho_s / (3 \rho_0) +
s^2 (I_0 - I_{-2})] / 2$, and $\eta = -I_0$. In our case of a pure
$s$-wave interaction, the Landau parameters in Eq. (68) of
\Refe{Leggett2} are given by $F_0 = 2 k_Fa/\pi$ and $F_1 = 0$. Then
the quantities $K_1$ and $Q$ of \Refe{Leggett2} correspond to our $\Pi$
and $\Pi_0$, respectively. As stated in \Refe{Leggett2}, the
long-wavelength and low-frequency limit is valid if $\hbar\omega, v_F
\hbar k \ll \Delta$. Since our semiclassical result coincides with
this limit of the quantum mechanical result, we conclude that this is
the condition for the validity of our semiclassical theory. A
calculation of the response function beyond the long-wavelength and
low-frequency limit can be found in \Refe{Minguzzi}.

\begin{figure}[t]
\includegraphics[width=6.5cm]{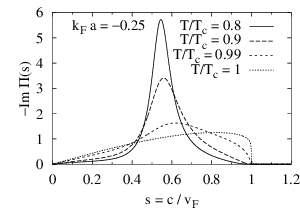}
\caption{Excitation spectrum $-\Im \Pi$ of a uniform medium as
function of the reduced sound velocity $s = c/v_F$ for different
temperatures.
\label{figsw}}
\end{figure}

The excitation spectrum of the system is characterized by the
imaginary part of $\Pi$, which is plotted in \Fig{figsw} for $k_F a =
-0.25$ and several temperatures between $0.8\, T_c$ and $T_c$. One can
see that at $0.8\, T_c$ the excitation spectrum exhibits a peak near
$s\approx 0.5$ which becomes broader and finally disappears when the
temperature approaches $T_c$.

In order to interprete this behavior, let us again consider the two
limits $T\to 0$ and $T\to T_c$. In the zero-temperature case, all
integrals containing the term $\fp(E_0)$ in the integrand vanish,
i.e., $\varphi = \rho_n/\rho_0 = I_n(s) = 0$, and the response
function reduces to
\begin{equation}
\Pi(s;0;k_F a) = \frac{1}{3s^2-1-2k_F a/\pi}\,.
\end{equation}
This means that the excitation spectrum is a $\delta$ function at
\begin{equation}
s = \sqrt{\frac{1}{3}+\frac{2k_F a}{3\pi}}\,, 
\end{equation}
corresponding to the hydrodynamic speed of sound.

In the other limit, $T\to T_c$, one has $\varphi = \rho_n/\rho_0 = 1$.
The integrals $I_n(s)$ reduce to
\begin{equation}
I_n(s;T/T_c\geq 1) = 1-s\arctanh\frac{1}{s}\,,
\end{equation}
independent of $n$, since the factors $\xi/E$ in the integrand of
\Eq{indef} can be replaced by $1$. As a consequence, the two equations
of the system (\ref{rho1phi1sw}) become identical and the coefficients
in front of $\phidh_1$ vanish, in accordance with the more general
arguments of \Sec{limitingcases}. Solving for $\rhoh_1$ gives
\begin{equation}
\Pi(s;T/T_c\geq 1;k_Fa) = \cfrac{-\Big(1-s\arctanh\cfrac{1}{s}\Big)}
  {1+\cfrac{2k_Fa}{\pi}\Big(1-s\arctanh\cfrac{1}{s}\Big)}\,,
\label{normalsw}
\end{equation}
in agreement with the usual result of Landau's Fermi-liquid theory for
the case of a pure $s$-wave interaction. If the interaction was
repulsive ($a > 0$), \Eq{normalsw} would have a pole at $s > 1$,
corresponding to the propagation of zero sound. However, here we are
considering the case of an attractive interaction, where zero sound
does not exist. Instead there is a continuous spectrum of
particle-hole excitations ranging from $s = 0$ to $s = 1$.

Our numerical results shown in \Fig{figsw} can be interpreted as
follows. At zero temperature, there exists a collective hydrodynamic
sound, which is undamped (at least within the present theoretical
treatment). As the temperature increases, a normal component
consisting of thermally excited quasiparticles builds up. However, at
temperatures where $\rho_n$ is already considerably different from
zero, the hydrodynamic sound is still practically undamped. The reason
for this is that all thermally excited quasiparticles contribute
equally to $\rho_n$, whereas only those quasiparticles whose velocity
$dE/dp \approx v_F \xi/E$ is at least equal to the sound velocity $c$
contribute to the Landau damping. At sufficiently high temperature,
the Landau damping becomes very strong and the hydrodynamic sound
ceases to exist. What remains is a continuum of particle-hole
excitations, and the interaction manifests itself only in the rounded
edge near $s = 1$.
\subsection{Quadrupole mode in a spherical trap}
\label{quadrupole}
Our main motivation for developing the present semiclassical approach
was to apply it to the case of trapped atomic Fermi gases. The
simplest example which comes to our mind is the quadrupole oscillation
of a Fermi gas in a spherical trap. Even in this case, the $\rv$
dependence of the equilibrium quantities induced by the trap potential
makes our equations very complicated. Since in this first
investigation we are interested in problems which can be solved
analytically, we will apply two additional simplifying approximations,
which allow us to obtain explicit solutions. A numerical method for
solving our equations without additional approximations will be
proposed at the end of this subsection.

Let us start with the linearized equation (\ref{qpvlasovlin}), which
has the form
\begin{equation}
\nud_1(\rv,\pv;t)-\poisson{E_0(\rv,\pv)}{\nu_1(\rv,\pv;t)}=F(\rv,\pv;t)\,.
\label{qpvlasovlinform}
\end{equation}
For its solution we adopt the Green function method used in
\Refe{KohlSchuck} to solve the linearized Vlasov equation for nuclear
giant resonances, which is formally very similar to our problem. The
starting point is to write the solution of \Eq{qpvlasovlinform} in the
form
\begin{multline}
\nu_1(\rv,\pv;t) = \int d\tp\int d^3 \rp \int d^3 \pp 
  G(\rv,\pv,\rvp,\pvp;t-\tp)\\
  \times F(\rvp,\pvp;\tp)\,,
\label{qpvlasovlinsol}
\end{multline}
where $G$ is the Green function of the differential operator on the
l.h.s. of \Eq{qpvlasovlinform}, satisfying
\begin{multline}
\Big[\frac{\partial}{\partial t}-\sum_{i=xyz}\Big(
  \frac{\partial E_0}{\partial r_i}\frac{\partial}{\partial p_i}
    -\frac{\partial E_0}{\partial p_i}\frac{\partial}{\partial r_i}\Big)
  \Big] G(\rv,\pv,\rvp,\pvp;t)\\
  = \delta(t)\delta(\rv-\rvp)\delta(\pv-\pvp)\,.
\label{qpgreen}
\end{multline}
Denoting by $\Rv(\rv,\pv;t)$ and $\Pv(\rv,\pv;t)$ the solutions of the
classical equations of motion
\begin{equation}
\dot{R}_i=\frac{\partial E_0(\Rv,\Pv)}{\partial P_i}\,,\qquad
\dot{P}_i=-\frac{\partial E_0(\Rv,\Pv)}{\partial R_i}
\label{classmotion}
\end{equation}
satisfying the initial conditions $\Rv(\rv,\pv;0) = \rv$ and
$\Pv(\rv,\pv;0) = \pv$, one can show that
\begin{equation}
G(\rv,\pv,\rvp,\pvp;t) = \theta(t)\delta[\rv-\Rv(\rvp,\pvp;t)]
  \delta[\pv-\Pv(\rvp,\pvp;t)]\,.
\end{equation}
fulfills \Eq{qpgreen}. Due to time-reversal symmetry and Liouville's
theorem, this Green function can be rewritten as
\begin{equation}
G(\rv,\pv,\rvp,\pvp;t) = \theta(t)\delta[\rvp-\Rv(\rv,-\pv;t)]
  \delta[\pvp+\Pv(\rv,-\pv;t)]\,.
\end{equation}
The latter form renders the phase-space integrals in
\Eq{qpvlasovlinsol} trivial. Changing the time integration variable
according to $\tau = t-\tp$, one obtains
\begin{equation}
\nu_1(\rv,\pv;t) = \int_0^\infty d\tau
  F[\Rv(\rv,-\pv;\tau),-\Pv(\rv,-\pv;\tau);t-\tau]\,.
\label{nu1tau}
\end{equation}
In the case of a harmonic perturbation,
\begin{equation}
V_{1\iext}(\rv;t) = \Vhat_{1\iext}(\rv) e^{-i\omega t}\,,
\end{equation}
the time dependence of $\nu_1$ as well as the explicit time dependence
of $F$ will be harmonic, too. We will denote the amplitudes by a hat
over the corresponding symbols. Multiplying \Eq{nu1tau} by $e^{i\omega
t}$, one finds that $\nuh_1$ is given by the Fourier integral
\begin{equation}
\nuh_1(\rv,\pv) = \int_0^\infty d\tau\,e^{i\omega \tau}
  \Fhat[\Rv(\rv,-\pv;\tau),-\Pv(\rv,-\pv;\tau)]\,.
\label{nu1omega}
\end{equation}

For the purpose of illustration we want to discuss a simple case in
which the classical trajectories are analytically known. We make two
approximations: First, we replace the $\rv$-dependent gap
$\Delta_0(\rv)$ by a constant $\Delta_0$. This approximation implies
that $\nu_{1\iev}$ is odd in $\xi$ and $\Deltat_{1\nu}$ can be
neglected, as it was the case in the preceding subsection. Second, we
will neglect effects from the Hartree mean-field as well in the
equilibrium state as in the deviations from equilibrium. The second
approximation, which is by far not as unrealistic as the first one,
amounts to neglecting all $g\rho_0$ and $g\rho_{1\nu}$ terms and
replacing the denominators $1+gA$ in \Eq{qpvlasovlin} by $1$. The trap
potential is assumed to be a spherical harmonic oscillator,
\begin{equation}
V_{0\iext} = \frac{1}{2}\,m\Omega^2 r^2\,.
\end{equation}
It is evident that the equations of motion (\ref{classmotion})
conserve $E_0$. However, if $\Delta_0$ is a constant, this implies
that $h_0$ is conserved, too, and the solutions of \Eqs{classmotion}
are closely related to the those of the ordinary harmonic
oscillator. Indeed, it is straight-forward to show that the
trajectories are given by
\begin{subequations}
\label{trajectories}
\begin{gather}
\Rv(\rv,\pv;t) = \rv \cos\frac{h_0\Omega t}{E_0}
  +\frac{\pv}{m\Omega}\sin\frac{h_0\Omega t}{E_0}\,,\\
\Pv(\rv,\pv;t) = \pv \cos\frac{h_0\Omega t}{E_0}
  -m\Omega\rv\sin\frac{h_0\Omega t}{E_0}\,.
\end{gather}
\end{subequations}
Since $h_0$ and $E_0$ are constants of the motion, they can likewise
be evaluated at $(\rv,\pv)$ or $(\Rv,\Pv)$.

Due to our approximations, the function $F$ [given by the r.h.s. of
\Eq{qpvlasovlin}] reduces to
\begin{multline}
\Fhat = -\fp(E_0) \Big[-\frac{h_0^2}{E_0^2}
  \frac{\pv}{m}\cdot\nablav(\Vhat_{1\iext}+i\hbar\omega\phih_1)\\
  +\frac{h_0}{E_0}\Big(\frac{\pv}{m}\cdot\nablav\Big)^2\hbar\phih_1
  -\Omega^2 \frac{h_0}{E_0}\rv\cdot\nablav\hbar\phih_1\Big]\,.
\label{fhat}
\end{multline}
As excitation we choose the quadrupole operator
\begin{equation}
\Vhat_{1\iext}(\rv) = \alpha m\Omega^2(\rv\otimes\rv)_{20}\,,
\label{v1q}
\end{equation}
where, explicitly,
\begin{equation}
(\vv\otimes\wv)_{20} = \sum_{\mu\nu} (1\mu1\nu|20) v_\mu w_\nu
  =\frac{2v_z w_z-v_x w_x-v_y w_y}{\sqrt{6}}\,.
\end{equation}
The prefactor in \Eq{v1q} has been chosen such that the coefficient
$\alpha$ is dimensionless. Due to the spherical symmetry of the trap,
the angular dependence of $\phih_1$ must be of the same quadrupolar
form as that of $\Vhat_{1\iext}$, but the radial dependence could in
principle be different. Here we make the ansatz that $\phih_1$ is
proportional to $\Vhat_{1\iext}$, i.e.,
\begin{equation}
\phih_1(\rv) = \beta\frac{m\Omega}{\hbar} (\rv\otimes\rv)_{20}\,,
\label{phi1qansatz}
\end{equation}
and we will show afterwards that with this ansatz for $\phih$ the
continuity equation can be satisfied by an appropriate choice of the
coefficient $\beta$. Quadratic ans\"atze like \Eq{phi1qansatz},
corresponding to a superfluid velocity field which is linear in the
coordinates, have frequently been used (see, e.g.,
\Refe{CozziniStringari}) for the calculation of the frequencies of
collective modes in the limit of superfluid hydrodynamics ($T = 0$).

Inserting \Eqs{v1q} and (\ref{phi1qansatz}) into \Eq{fhat} and using
the explicit form of the trajectories, \Eq{trajectories}, we can
evaluate the Fourier integral in \Eq{nu1omega}, with the result
\begin{widetext}
\begin{equation}
\nuh_1 = -2\fp(E_0)\left[
\Big(\frac{(\pv\otimes\pv)_{20}}{m}
    -m\Omega^2(\rv\otimes\rv)_{20}\Big)\frac{h_0}{E_0}\,
  \cfrac{i\beta\cfrac{\omega}{\Omega}\,\cfrac{\Delta_0^2}{E_0^2}
    -\alpha\cfrac{h_0^2}{E_0^2}}
    {\cfrac{\omega^2}{\Omega^2}-4\cfrac{h_0^2}{E_0^2}}
+
\Omega(\rv\otimes \pv)_{20}\frac{h_0^2}{E_0^2}\left(\beta
  -\cfrac{4\beta\cfrac{\Delta_0^2}{E_0^2}+i\alpha\cfrac{\omega}{\Omega}}
    {\cfrac{\omega^2}{\Omega^2}-4\cfrac{h_0^2}{E_0^2}}\right)
\right]\,,
\end{equation}
\end{widetext}
Now we have to calculate the corresponding current $\jvh$ and density
oscillations $\rhoh_{1\nu}$. As detailed in the preceding subsection,
this is accomplished by integrating $\pv \nuh_1/m$ and $\nuh_1$,
respectively, over $\pv$. Replacing in the integrals $p^2dp$ by
$mp_F(\rv)d\xi$, $p$ by $p_F(\rv)$, $h_0$ by $\xi$, and $E_0$ by
$\sqrt{\xi^2+\Delta_0^2}$, we obtain
\begin{gather}
\jvh_{1\nu}(\rv) = \rho_0(\rv)\Omega
  \big(\beta[\varphi-4I_{22}(z)]-i\alpha zI_{20}(z)\big)\,
  \nablav(\rv\otimes\rv)_{20}\,,\\
\rhoh_{1\nu}(\rv) = \frac{m^2\Omega^2 p_F(\rv)}{\pi^2\hbar^3}
  [\alpha I_{40}(z)-i\beta zI_{22}(z)]
    (\rv\otimes\rv)_{20}\,,
\end{gather}
with the abbreviations
\begin{gather}
z = \frac{\omega}{\Omega}\,,\\
I_{ij}(z) = -\int d\xi f^\prime(E)\frac{\xi^i\Delta_0^j}{E^{i+j}}\,
  \frac{1}{z^2-4\xi^2/E^2}\,,
\label{iijdef}
\end{gather}
where $E = \sqrt{\xi^2+\Delta_0^2}$. From its definition it is evident
that $I_{40} = I_{20}-I_{22}$, such that it is sufficient to evaluate
two of these integrals numerically. The functions $I_{ij}(z)$ have a
branch cut along the real axis from $z = -2$ to $z = 2$. Remember that
$\omega$, and therefore also $z$, is assumed to have an infinitesimal
positive imaginary part, fixing the sign of the imaginary part of
$I_{ij}(z)$.

As stated above, the coefficient $\beta$ must be determined by the
continuity equation (\ref{contilin}). Due to our approximation to
neglect the Hartree field, the denominator $1+gA(\rv)$ in the first
term of \Eq{contilin} can be replaced by $1$, and the Fermi momentum
can be given in closed form:
\begin{equation}
p_F(\rv) = \sqrt{2m\Big(\mu-\frac{1}{2}m\Omega^2 r^2\Big)}\,.
\end{equation}
Inserting the results for $\jvh_{1\nu}$ and $\rhoh_{1\nu}$ into the
continuity equation, one finds that the ansatz (\ref{phi1qansatz})
indeed allows to satisfy the continuity equation, and the
corresponding solution for the coefficient $\beta$ reads
\begin{equation}
\beta = iz\alpha\frac{1-\varphi+2 I_{22}(z)}
    {(1-\varphi)(z^2-2)+2I_{22}(z)(z^2-4)}\,.
\end{equation}
This expression can be used to obtain $\rhoh_{1\nu}$. Here we will
immediately give the result for the amplitude of the total density
oscillations, i.e., $\rhoh_1 = \rhoh_{1\nu}-A
(\Vhat_{1\iext}+i\hbar\omega\phih_1)$, which we write in the form
\begin{equation}
\rhoh_1(\rv) = \alpha\frac{m^2\Omega^2p_F(\rv)}{\pi^2\hbar^3}
  (\rv\otimes\rv)_{20} \Pi(z)
\end{equation}
with
\begin{equation}
\Pi(z) = I_{20}(z)
  +\frac{[1-\varphi+2I_{22}(z)][1-\varphi+4I_{22}(z)]}
     {(1-\varphi)(z^2-2)+2 I_{22}(z)(z^2-4)}\,.
\end{equation}

Before discussing numerical results, let us again study the two
extreme cases $T=0$ and $T\geq T_c$. In the zero-temperature limit,
all integrals $\varphi$ and $I_{ij}$ are zero, and hence the response
function becomes
\begin{equation}
\Pi(z;T/T_c=0) = \frac{1}{z^2-2}\,,
\end{equation}
i.e., it has a single pole at the hydrodynamical frequency
\begin{equation}
\omega = \sqrt{2}\Omega\,.
\end{equation}
In the case $T \geq T_c$, i.e., in the normal phase, we have $\varphi
= 1$, and in the definition (\ref{iijdef}) we can replace $\Delta_0$
and $E_0$ by $0$ and $\xi$, respectively, such that we obtain
\begin{subequations}
\begin{gather}
I_{20}(z;T/T_c\geq 1) = \frac{1}{z^2-4}\,,\\
  I_{22}(z;T/T_c\geq 1) = 0\,.
\end{gather}
\end{subequations}
Thus the response function reduces to
\begin{equation}
\Pi(z;T/T_c\geq 1) = \frac{1}{z^2-4}\,.
\end{equation}
Like in the zero-temperature case, we have a single pole, but now at a
frequency which is higher by a factor of $\sqrt{2}$. The reason for
the difference of the two frequencies is as follows. In the superfluid
phase, the momentum distribution stays spherical during the
oscillation. In contrast to this, in the normal phase, the momentum
distribution is deformed in the opposite direction as the density in
coordinate space. This deformation of the Fermi sphere costs kinetic
energy, which increases the restoring force and thereby the frequency
of the oscillation.

\begin{figure}[t]
\includegraphics[width=6.5cm]{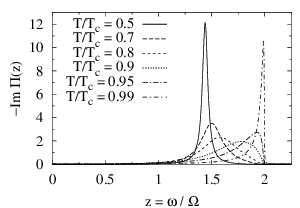}
\caption{Quadrupole excitation spectrum $-\Im \Pi$ of a harmonically
trapped gas as a function of the excitation frequency (in units of the
trap frequency) for different temperatures. The spatial dependence of
the gap as well as the Hartree mean-field have been neglected.
\label{figsq}}
\end{figure}

At intermediate temperatures $0<T<T_c$, the excitation spectrum is
continuous and it is characterized by the imaginary part of $\Pi(z)$,
which is shown in \Fig{figsq} for a set of temperatures between $0.5\,
T_c$ and $T_c$. At $0.5\, T_c$, the spectrum exhibits a sharp peak at
the hydrodynamic frequency $z = \sqrt{2}$, the weak broadening being
due to Landau damping. With increasing temperature, the Landau damping
becomes more important, and at the same time the centroid of the
distribution moves to higher frequencies. Above $0.8\, T_c$, however,
the width of the peak does not increase any more with temperature, but
it decreases. Finally, when the temperature approaches $T_c$, the peak
becomes again very sharp and, not surprisingly, it lies at the
frequency $z = 2$ predicted by the Vlasov equation.

One might ask the question whether the approximations made in this
subsection are justified or not. Let us therefore compare our results
with those of a QRPA (quasiparticle random-phase approximation)
calculation \cite{GrassoKhan}, where, apart from the mean-field
approximation leading to the Bogoliubov-de Gennes equations, no
approximations are made. Qualitatively our semiclassical results show
the main features of this quantum-mechanical calculation: the
hydrodynamic mode at zero temperature, its damping at intermediate
temperatures, and the subsequent reappearance of an undamped
collective mode with a higher frequency in the normal phase. That our
frequency in the normal phase is exactly equal to $z=2$ is a
consequence of neglecting the Hartree mean field. However, in the
range of validity of our theory ($k_F|a|\ll 1$), the Hartree mean
field cannot shift the frequency very much (in \Refe{GrassoKhan}, e.g.,
the frequency is shifted from $2$ to $\approx 2.2$). We therefore
believe that this effect is not very important. More problematic is
the constant-gap approximation which we needed for the analytical
solution of the equations of motion (\ref{classmotion}). Because of
this approximation, there are no quasiparticles having energies below
$\Delta_0$, and as a consequence, the Landau damping sets in at rather
high temperatures. In the full calculation, however, the lowest-lying
quasiparticles are those whose wave functions are localized near the
surface, where the gap is small, and which have much smaller energies
than the central value of the gap. Therefore within the full
calculation the Landau damping is already quite important at very low
temperatures. In the semiclassical formalism these low-lying
quasiparticles can be understood as quasiparticles bouncing back and
forth between the potential wells created by the trap potential and
the spatially varying gap (Andreev reflection)
\cite{OhashiGriffin,BruunHeiselberg}. The inclusion of this effect
would require a numerical solution of the equations of motion
(\ref{classmotion}).

This leads us to a possible numerical method for solving even the
original (i.e., not linearized) kinetic equation (\ref{qpvlasov}). In
nuclear physics, the Vlasov equation (usually complemented by a
collision term) is routinely solved by the so-called test-particle
method, e.g., in order to simulate heavy-ion collisions
\cite{BertschGupta}. Recently this method has also been applied to the
solution of the Vlasov equation with collision term for trapped atomic
Fermi gases \cite{Toschi} and of a Vlasov-like equation for trapped
fermion-boson mixtures \cite{Maruyama}, and it seems to be
straight-forward to generalize it to our case. The basic idea of the
method is as follows. Instead of calculating the time evolution of the
continuous quasiparticle distribution function $\nu(\rv,\pv;t)$, one
can use a finite number of ``test-quasiparticles'' and follow their
motion in phase-space by solving numerically the equations of motion
\begin{equation}
\dot{R}_i=\frac{\partial E(\Rv,\Pv)}{\partial P_i}\,,\qquad
\dot{P}_i=-\frac{\partial E(\Rv,\Pv)}{\partial R_i}
\end{equation}
for all test-quasiparticles simultaneously. Of course, the
quasiparticle energy $E(\rv,\pv;t)$ contains the mean-fields
$g\rho(\rv;t)$ and $\Deltat(\rv;t)$, which must be calculated at each
time step from the actual quasiparticle distribution. At the same
time, the phase $\phi$ must be calculated at each time step from the
continuity equation. This seems to be a tractable task, which will be
adressed in a subsequent publication.

\section{Summary and conclusions}
\label{summary}
In the first part of the present article, we derived a set of
semiclassical equations describing the dynamics of a collisionless
superfluid Fermi gas by taking the $\hbar\to 0$ limit of the TDHFB or
time-dependent Bogoliubov-de Gennes equations. In the limits of zero
and critical temperature, these equations reproduce superfluid
hydrodynamics and the Vlasov equation, respectively. At intermediate
temperatures, there is a complicated interplay between the dynamics of
the superfluid component of the system, governed by the function
$\phi(\rv)$ which is related to the phase of the gap $\Delta(\rv)$,
and the dynamics of the normal component, which is described by the
quasiparticle distribution function $\nu(\rv,\pv)$. The dynamical
equation for $\nu$ formally corresponds to the usual Vlasov equation
with $\varrho$ and $h$ replaced by $\nu$ and $E$, respectively, while
the function $\phi$ is determined by the continuity equation. The
latter point can be seen most easily in the linearized version of the
equations for small deviations from equilibrium.

In the second part, we gave an illustration of our equations by
applying them to two simple cases, where analytical solutions could be
found. The first example we studied was a sound wave traveling in a
uniform system. In this case we could reproduce the usual hydrodynamic
speed of sound at zero temperature. At non-zero temperatures below
$T_c$, the sound wave suffers strong Landau damping because of its
coupling to thermally excited quasiparticles. For $T\to T_c$ the
excitation spectrum continuously goes over into that of the usual
particle-hole continuum (with RPA corrections) which is found above
$T_c$.

The second example was the quadrupole mode of a Fermi gas in a
spherical trap. Applying the approximation of a constant gap and
neglecting the Hartree field, we were able to solve the linearized
quasiparticle kinetic equation exactly also for this case. We could
qualitatively reproduce the most important results of quantum
mechanical QRPA calculations: At zero temperature, there is an
undamped collective mode at the hydrodynamic frequency $\omega =
\sqrt{2}\Omega$, which becomes strongly damped at low temperatures
($0<T \ll T_c$). At a certain temperature the damping rate reaches a
maximum, and above that temperature it decreases until at $T = T_c$ an
undamped collective mode reappears at the frequency predicted by the
Vlasov equation, which is higher than the hydrodynamic
frequency. However, quantitatively the agreement with the QRPA
calculation is not yet satisfactory, essentially because we replaced
the gap by an $\rv$-independent constant. We suggested to use the
test-particle method for the numerical solution of the equations in
the case of a spatially varying gap $\Delta(\rv)$, which at the same
time would allow to include the Hartree field and to treat strong
deviations from equilibrium, like the expansion of the gas after the
trap is switched off.

As long as the gas is close to equilibrium, i.e., as long as the Fermi
surface can be regarded as spherical, the effect of collisions is very
small due to Pauli blocking of the final states. However, in the case
of strong deviations from equilibrium, like during the expansion of
the cloud when the trap is switched off, the deformation of the Fermi
sphere can lead to rather important collisional effects
\cite{JacksonPedri}. Therefore, in this case it is necessary to
include the collision term into the theory. This is an interesting
problem which should be addressed in a future investigation. For
normal-fluid trapped Fermi gases there exist already some calculations
which take the collision term into account
\cite{Toschi,JacksonPedri}. The more complicated case of paired Fermi
systems with collisions has been considered, e.g., in the context of
superfluid $^3$He \cite{SereneRainer,Woelfle}.

Since our equations were obtained as the $\hbar\to 0$ limit of the
TDHFB equations, only the leading gradient terms are included. This
can be seen, e.g., in our results for the sound wave, which in fact
are the long-wavelength and low-frequency limit of the quantum
mechanical result which can be obtained by diagrammatic
techniques. For the system in a trap this means in particular that the
gradients of the trapping potential, which are proportional to the
trap frequency $\Omega$, must not be too strong. But what is ``too
strong''? In the normal phase ($\Delta = 0$) the $\hbar\to 0$ limit
works extremely well if $\hbar\Omega\ll\mu$, which is always the case
in the experimental situations. In the superfluid phase ($\Delta\neq
0$) the relevant condition reads $\hbar\Omega\ll\Delta$
\cite{UrbanSchuck,GrassoKhan}, which is much more difficult to
satisfy, especially for the radial trap frequency which is usually
much larger than the axial one. In spite of this limitation, we
believe that the semiclassical approach presented here will be useful,
since at present no quantum mechanical calculation is able to describe
the dynamics of systems with more than $\approx 10^4$ atoms,
especially in the case of deformed traps.

\begin{acknowledgments}
We are grateful to E. Balbutsev for numerous discussions. P.S. wants
to thank F. Gulminelli for an earlier collaboration on dynamical aspects
of pairing.
\end{acknowledgments}

\end{document}